\documentclass[twocolumn,showpacs,preprintnumbers]{revtex4}%
\usepackage{amssymb}
\usepackage{amsmath}
\usepackage{graphicx}
\usepackage{epstopdf}
\usepackage{dcolumn}
\usepackage{bm}
\usepackage{amsfonts}%
\UseRawInputEncoding
\begin{document}
\title{Quantum measurement with recycled photons}
\author{Eyal Buks}
\affiliation{Andrew and Erna Viterbi Department of Electrical Engineering, Technion, Haifa
32000 Israel}
\author{Banoj Kumar Nayak}
\affiliation{Andrew and Erna Viterbi Department of Electrical Engineering, Technion, Haifa
32000 Israel}
\date{\today }

\begin{abstract}
We study a device composed of an optical interferometer integrated with a
ferri-magnetic sphere resonator (FSR). Magneto-optic coupling can be employed
in such a device to manipulate entanglement between optical pulses that are
injected into the interferometer and the FSR. The device is designed to allow
measuring the lifetime of such macroscopic entangled states in the region
where environmental decoherence is negligibly small. This is achieved by
recycling the photons interacting with the FSR in order to eliminate the
entanglement before a pulse exits the interferometer. The proposed experiment
may provide some insight on the quantum to classical transition associated
with a measurement process.

\end{abstract}
\maketitle



%It is always \today, today,
%but any date may be explicitly specified

%PACS, the Physics and Astronomy
%Classification Scheme.

\section{Introduction}

Consider two successive quantum measurements \cite{Johansen_5760}. In the
first one, which is performed at time $t_{1}$, the observable $A_{1}$ is being
measured, whereas in the second one, which is performed at a later time
$t_{2}\geq t_{1}$, the observable $A_{2}$ is being measured. Let
$\mathcal{A}_{1}$ ($\mathcal{A}_{2}$) be the outcome of the first (second)
measurement, and $\left\{  a_{n,k}\right\}  _{k}$ be the set of eigenvalues of
the observable $A_{n}$, where $n\in\left\{  1,2\right\}  $. The probability
that the measurement at time $t_{2}$ of the observable $A_{2}$ yields the
value $a_{2,k_{2}}$, namely, the probability that $\mathcal{A}_{2}=a_{2,k_{2}%
}$, is denoted by $p_{2}\left(  k_{2}\right)  $ Two methods for the
calculation of $p_{2}\left(  k_{2}\right)  $ are considered below. In the
first one, the time evolution from an initial time $t_{0}<t_{1}$ to time
$t_{2}$ is assumed to be purely unitary, and the probability $p_{2}\left(
k_{2}\right)  $ for the measurement at time $t_{2}$ is calculated using the
Born rule. The second method is based on the assumption that the unitary
evolution is disturbed at time $t_{1}$, at which the density operator of the
system undergoes a collapse
\cite{Schrodinger_807,Legget_R415,Aharonov_359,Mermin_38,Mooij_401,Bell_208,Zurek_1516}
corresponding to the measurement of the observable $A_{1}$. Note that for both
methods the coupling between the quantum subsystem and its measuring apparatus
is taken into account in the unitary time evolution
\cite{von_Neumann_Mathematical_Foundations,Aharonov_11,Peres_book,Braginsky_Quantum_Meas,Ruskov_200404}%
. Under what conditions the probability $p_{2}\left(  k_{2}\right)  $ is
affected \cite{Legget_857} by whether a collapse has occurred, or has not
occurred, at the earlier time $t_{1}$?

A sufficient condition, which ensures that the collapse at time $t_{1}$ has no
effect on the probability $p_{2}\left(  k_{2}\right)  $, is discussed below.
This sufficient condition can be expressed as $\left[  A_{2}\left(
t_{2}\right)  ,A_{1}\left(  t_{1}\right)  \right]  =0$, where $A_{1}\left(
t_{1}\right)  $ and $A_{2}\left(  t_{2}\right)  $ are the Heisenberg
representations of the $A_{1}$ and $A_{2}$ operators, respectively [see Eq.
(8.501) of \cite{Buks_QMLN}]. As is explained below, this condition is
satisfied for the vast majority of experimental setups used to study quantum systems.

Commonly the entire system can be composed into a quantum subsystem (QS) under
study, and one or more ancilla subsystems (AS) that are used for probing the
QS. Moreover, very commonly, the process of measurement is based on scattering
of AS particles (electrons, photons, phonons, magnons, etc.) by the QS under
study. In such a scattering process, the QS is bombarded by incoming AS
particles. Properties of the QS are inferred from measured properties of the
scattered AS particles. For this type of measurements the observables $A_{1}$
and $A_{2}$ are operators of the AS, and are independent on the degrees of
freedom of the QS.

For the above-discussed two successive measurements of a given QS, two cases
are considered below. For the first one, which is the common case, the ancilla
particles that are used for the first measurement are not used for the second
one. The two independent ASs associated with the two successive measurements
are denoted by AS1 and AS2, respectively. For this case the observable $A_{1}$
($A_{2}$) is an operator of AS1 (AS2), and consequently the condition $\left[
A_{2}\left(  t_{2}\right)  ,A_{1}\left(  t_{1}\right)  \right]  =0$ is
satisfied, therefore, any collapse-induced effect on the probability
$p_{2}\left(  k_{2}\right)  $ corresponding to the second measurement is excluded.

For the second case, AS particles used for performing the first measurement
are recycled in order to participate in the second measurement as well. For
this case, which is far less common, the condition $\left[  A_{2}\left(
t_{2}\right)  ,A_{1}\left(  t_{1}\right)  \right]  =0$ can be violated, and
consequently collapse-induced effect on $p_{2}\left(  k_{2}\right)  $ cannot
be ruled out. The possibility that the condition $\left[  A_{2}\left(
t_{2}\right)  ,A_{1}\left(  t_{1}\right)  \right]  =0$ is violated raises some
concerns regarding the mathematical self-consistency of quantum mechanics
\cite{Penrose_4864,Leggett_939,Leggett_022001} (note that this is unrelated to
compatibility with the principle of causality).

\section{Optical interferometer}

\begin{figure}
[ptb]
\begin{center}
\includegraphics[
trim=0.008474in 0.000000in -0.008474in 0.000000in,
height=0.9046in,
width=3.2396in
]%
{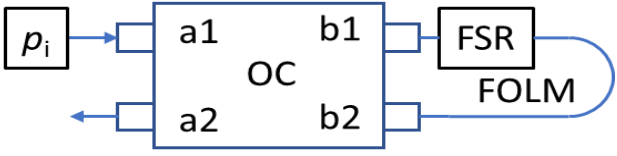}%
\caption{FOLM interferometer. Light is injected into port a1 of the OC, and
detection is performed using a PD connected to port a2. The FSR is integrated
inside a microwave cavity \cite{Mathai_054428} (not shown in the sketch).}%
\label{FigFOLM}%
\end{center}
\end{figure}

In the proposed experimental setup, a fiber optical loop mirror (FOLM)
\cite{Mortimore_1217,Ibarra_191} is employed in order to allow performing
measurements with recycled photons (see Fig. \ref{FigFOLM}). A short optical
pulse having state of polarization (SOP) $\left\vert p_{\mathrm{i}%
}\right\rangle $ is injected into port a1 of an optical coupler (OC). A
Ferrimagnetic sphere resonator (FSR)
\cite{kittel1976introduction,Chai_1900252} is integrated into the fiber loop
of the FOLM near port b1 of the OC. Magneto-optic (MO) coupling
\cite{Freiser_152,Pershan_1482} between the optical pulse and the FSR gives
rise to both the Faraday-Voigt effect, which accounts for the change in the
optical SOP, and the inverse Faraday effect (IFE)
\cite{Braggio_107205,Colautti_thesis,Kirilyuk_2731,Kirilyuk_026501,Kimel_655,VanderZiel_190,Pershan_574,Hansteen_014421,Kirilyuk_748}%
, which accounts for the optically-induced change in the FSR state of
magnetization (SOM). The externally injected optical pulse interacts with the
FSR at times $t_{1}$, and $t_{2}>t_{1}$, and the experimental setup allows the
violation of the condition $\left[  A_{2}\left(  t_{2}\right)  ,A_{1}\left(
t_{1}\right)  \right]  =0$, where $A_{1}\left(  t_{1}\right)  $ and
$A_{2}\left(  t_{2}\right)  $ are the corresponding observables. The time
difference $t_{2}-t_{1}$ is set by adjusting the length of the fiber loop
(labelled as FOLM in Fig. \ref{FigFOLM}). The transmitted signal at port a2 of
the OC is measured using a photodetector (PD).%

%EndExpansion

The OC is characterized by forward (backward) transmission $t$ ($t^{\prime}$)
and reflection $r$ ($r^{\prime}$) amplitudes. Incoming amplitudes $\bar
{E}_{\mathrm{in}}=\left(
\begin{array}
[c]{cccc}%
E_{_{\rightarrow}}^{a_{1}} & E_{_{\rightarrow}}^{a_{2}} & E_{_{\leftarrow}%
}^{b_{1}} & E_{\leftarrow}^{b_{2}}%
\end{array}
\right)  ^{\mathrm{T}}$ are related to outgoing amplitudes $\bar
{E}_{\mathrm{out}}=\left(
\begin{array}
[c]{cccc}%
E_{_{\leftarrow}}^{a_{1}} & E_{_{\leftarrow}}^{a_{2}} & E_{_{\rightarrow}%
}^{b_{1}} & E_{\rightarrow}^{b_{2}}%
\end{array}
\right)  ^{\mathrm{T}}$ by $\bar{E}_{\mathrm{out}}=S\bar{E}_{\mathrm{in}}$
(subscript horizontal arrow indicates propagation direction, and superscripts
indicates OC port label), where the scattering matrix $S$ is given by (it is
assumed that all scattering coefficients are polarization independent)%
\begin{equation}
S=\left(
\begin{array}
[c]{cccc}%
0 & 0 & t^{\prime} & r^{\prime}\\
0 & 0 & r^{\prime} & t^{\prime}\\
t & r & 0 & 0\\
r & t & 0 & 0
\end{array}
\right)  \;.
\end{equation}
Unitarity $S^{\dag}S=1$ implies that $\left\vert t\right\vert ^{2}+\left\vert
r\right\vert ^{2}=\left\vert t^{\prime}\right\vert ^{2}+\left\vert r^{\prime
}\right\vert ^{2}=1$ and $\operatorname{Re}\left(  r^{\ast}t\right)
=\operatorname{Re}\left(  r^{\prime\ast}t^{\prime}\right)  =0$. Time reversal
symmetry $S^{\mathrm{T}}=S$ implies that $t^{\prime}=t$ and $r^{\prime
}=r=it\left\vert r/t\right\vert $.

The transmission (reflection) coefficient $t$ ($r$) is the amplitude of the
sub-pulse circulating the FOLM in the clockwise (counter clockwise) direction.
The MO coupling gives rise to a change in both the optical SOP and the FSR
SOM. These states for the clockwise (counter clockwise) direction are labelled
by $\left\vert p_{+}\right\rangle _{\mathrm{P}}$ and $\left\vert
m_{+}\right\rangle _{\mathrm{M}}$ ($\left\vert p_{-}\right\rangle
_{\mathrm{P}}$ and $\left\vert m_{-}\right\rangle _{\mathrm{M}}$),
respectively (note that these states, which are allowed to change in time, are
assumed to be normalized). The state vector $\left\vert \psi_{\mathrm{f}%
}\right\rangle $, which represents a final state after the pulse has left the
interferometer, can be expressed as%
\begin{align}
\left\vert \psi_{\mathrm{f}}\right\rangle  &  =tr^{\prime}\left\vert
a_{1}\longleftarrow,p_{+},m_{+}\right\rangle +rt^{\prime}\left\vert
a_{1}\longleftarrow,p_{-},m_{-}\right\rangle \nonumber\\
&  +tt^{\prime}\left\vert a_{2}\longleftarrow,p_{+},m_{+}\right\rangle
+rr^{\prime}\left\vert a_{2}\longleftarrow,p_{-},m_{-}\right\rangle
\;,\nonumber\\
&  \label{|psi_f>}%
\end{align}
where $\left\vert \mathrm{T},p,m\right\rangle =\left\vert \mathrm{T}%
\right\rangle _{\mathrm{I}}\otimes\left\vert p\right\rangle _{\mathrm{P}%
}\otimes\left\vert m\right\rangle _{\mathrm{M}}$ denotes a state having pulse
in interferometer port $\mathrm{T}$, optical polarization $p$, and FSR
magnetization $m$.

Let $\left\{  \left\vert p_{n^{\prime}}\right\rangle _{\mathrm{P}}\right\}  $
($\left\{  \left\vert m_{n^{\prime\prime}}\right\rangle _{\mathrm{M}}\right\}
$) be an orthonormal basis for the Hilbert space of optical SOP (FSR SOM). The
transmission $p_{\mathrm{T}}$ and reflection $p_{\mathrm{R}}$ probabilities
are found by tracing out%
\begin{align}
p_{\mathrm{T}}  &  =\sum_{n^{\prime},n^{\prime\prime}}\left\vert \left\langle
\psi_{\mathrm{f}}\right\vert \left(  \left\vert a_{2}\longleftarrow
\right\rangle _{\mathrm{I}}\otimes\left\vert p_{n^{\prime}}\right\rangle
_{\mathrm{P}}\otimes\left\vert m_{n^{\prime\prime}}\right\rangle _{\mathrm{M}%
}\right)  \right\vert ^{2}\;,\\
p_{\mathrm{R}}  &  =\sum_{n^{\prime},n^{\prime\prime}}\left\vert \left\langle
\psi_{\mathrm{f}}\right\vert \left(  \left\vert a_{1}\longleftarrow
\right\rangle _{\mathrm{I}}\otimes\left\vert p_{n^{\prime}}\right\rangle
_{\mathrm{P}}\otimes\left\vert m_{n^{\prime\prime}}\right\rangle _{\mathrm{M}%
}\right)  \right\vert ^{2}\;,
\end{align}
hence (note that $\sum_{n^{\prime}}\left\vert p_{n^{\prime}}\right\rangle
_{_{\mathrm{PP}}}\left\langle p_{n^{\prime}}\right\vert =1_{_{\mathrm{P}}}$,
$\sum_{n^{\prime\prime}}\left\vert m_{n^{\prime\prime}}\right\rangle
_{\mathrm{MM}}\left\langle m_{n^{\prime\prime}}\right\vert =1_{_{\mathrm{M}}}%
$, and recall that $\left\vert p_{\pm}\right\rangle _{\mathrm{P}}$ and
$\left\vert m_{\pm}\right\rangle _{\mathrm{M}}$ are normalized, and that
$t^{\prime}=t$ and $r^{\prime}=r=it\left\vert r/t\right\vert $)%
\begin{align}
p_{\mathrm{T}}  &  =\left(  \left\vert t\right\vert ^{2}-\left\vert
r\right\vert ^{2}\right)  ^{2}+4\left\vert tr\right\vert ^{2}\eta
\;,\label{p_T}\\
p_{\mathrm{R}}  &  =4\left\vert tr\right\vert ^{2}\left(  1-\eta\right)  \;,
\label{p_R}%
\end{align}
where%
\begin{equation}
\eta=\frac{1-\operatorname{Re}\left(  \chi_{\mathrm{P}}\chi_{\mathrm{M}%
}\right)  }{2}\;, \label{eta}%
\end{equation}
and where $\left.  \chi_{\mathrm{P}}=\right.  _{\mathrm{P}}\left\langle
p_{+}\right.  \left\vert p_{-}\right\rangle _{\mathrm{P}}$ and $\left.
\chi_{\mathrm{M}}=\right.  _{\mathrm{M}}\left\langle m_{+}\right.  \left\vert
m_{-}\right\rangle _{\mathrm{M}}$. Note that $p_{\mathrm{T}}+p_{\mathrm{R}}%
=1$\ (recall that $\left\vert t\right\vert ^{2}+\left\vert r\right\vert
^{2}=1$). In the absence of any MO coupling, i.e. when $\chi_{\mathrm{P}}%
\chi_{\mathrm{M}}=1$, $\eta=0$, whereas $\eta=1/2$ for the opposite extreme
case of $\chi_{\mathrm{P}}\chi_{\mathrm{M}}=0$. For the case of a 3dB OC (i.e.
when $\left\vert t\right\vert ^{2}=\left\vert r\right\vert ^{2}=1/2$) this
becomes $p_{\mathrm{T}}=\eta$ and $p_{\mathrm{R}}=1-\eta$. Thus, in the
absence of any MO coupling and for a 3dB OC the transmission probability
$p_{\mathrm{T}}$ vanishes. This unique property, which originates from
destructive interference in the FOLM interferometer, allows sensitive
measurement of the effect of MO coupling.

The parameter $\chi_{\mathrm{P}}$ characterizes the change in SOP induced by
the Faraday-Voigt effect, whereas the change in the FSR SOM induced by the IFE
\cite{Crescini_1,Braggio_107205,Hisatomi_174427} is characterized by the
parameter $\chi_{\mathrm{M}}$. Both effects originate from the MO coupling
between the optical pulses and the FSR, and the Verdet constant
\cite{VanderZiel_190,Donati_372,Freiser_152,Pershan_1482} is proportional to
both induced changes in SOP and SOM \cite{Battiato_014413} [see also Eq.
(2.316) of \cite{Buks_WPLN}]. Based on appendix \ref{App_MO}, which reviews MO
coupling, the parameter $\eta$ is estimated.

Two configurations are considered below. For the first one $\mathbf{\hat{q}%
}\parallel\mathbf{H}_{\mathrm{dc}}$, whereas $\mathbf{\hat{q}}\perp
\mathbf{H}_{\mathrm{dc}}$ for the second configuration, where $\mathbf{\hat
{q}}$ is a unit vector parallel to the optical propagation direction, and
where $\mathbf{H}_{\mathrm{dc}}$ is the static magnetic field externally
applied to the FSR. The angular frequency of the Kittel mode $\omega
_{\mathrm{m}}$ is related to $H_{\mathrm{dc}}$ by $\omega_{\mathrm{m}}%
=\gamma_{\mathrm{e}}\mu_{0}H_{\mathrm{dc}}$, where $\gamma_{\mathrm{e}}%
/2\pi=28%
%TCIMACRO{\unit{GHz}}%
%BeginExpansion
\operatorname{GHz}%
%EndExpansion%
%TCIMACRO{\unit{T}}%
%BeginExpansion
\operatorname{T}%
%EndExpansion
^{-1}$\ is the gyromagnetic ratio, and $\mu_{0}$ is the free space
permeability (magnetic anisotropy is disregarded). For both cases it is shown
below that, on one hand, the intermediate value of $\operatorname{Re}\left(
\chi_{\mathrm{P}}\chi_{\mathrm{M}}\right)  $ during the time interval $\left[
t_{1},t_{2}\right]  $ can be made significantly smaller than unity, whereas,
on the other hand, the final (i.e. after time $t_{2}$) value of
$\operatorname{Re}\left(  \chi_{\mathrm{P}}\chi_{\mathrm{M}}\right)  $ can be
made very close to unity [see Eq. (\ref{eta})]. Hence, for these cases the
transmitted signal at port a2 is strongly affected by the level of unitarity
in the time evolution of the system prior to time $t_{2}$.

The change in SOP for the first configuration is dominated by the Faraday
effect, whereas the Voigt effect, which is much weaker [see Eqs. (\ref{k_CB}),
(\ref{k_LB}) and (\ref{J sphere}) of appendix \ref{App_MO}, and note that
$Q_{\mathrm{s}}\ll1$] accounts for the change in SOP for the second
configuration. In the analysis below, the change in SOP is disregarded for the
second configuration (i.e. it is assumed that $\chi_{\mathrm{P}}=1$).

The IFE gives rise to an effective magnetic field $\mathbf{H}_{\mathrm{IFE}}$,
which is parallel to the optical propagation direction $\mathbf{\hat{q}}$, and
it has a magnitude proportional to $I_{\mathrm{p+}}-I_{\mathrm{p-}}$, where
$I_{\mathrm{p+}}$ ($I_{\mathrm{p-}}$) is the optical energy carried by
right-hand $\left\vert R\right\rangle $ (left-hand $\left\vert L\right\rangle
$) circular SOP \cite{VanderZiel_190} [see Eq. (\ref{H_IFE}) of appendix
\ref{App_MO}]. With femtosecond optical pulses this optically-induced magnetic
field $\mathbf{H}_{\mathrm{IFE}}$ can be employed for ultrafast manipulation
of the SOM \cite{Kimel_275,Juraschek_094407,Juraschek_043035}. For the first
configuration (for which $\mathbf{\hat{q}}\parallel\mathbf{H}_{\mathrm{dc}}$),
it is expected that the change in the SOM due to the IFE will be relatively
small (since $\mathbf{H}_{\mathrm{IFE}}\parallel\mathbf{H}_{\mathrm{dc}}$, and
the magnetization is assumed to be nearly parallel to $\mathbf{H}%
_{\mathrm{dc}}$). In the analysis below, the change in SOM is disregarded for
the first configuration (i.e. it is assumed that $\chi_{\mathrm{M}}=1$). For
the second configuration (for which $\mathbf{\hat{q}}\perp\mathbf{H}%
_{\mathrm{dc}}$), on the other hand, the IFE gives rise to a much larger
effect (since $\mathbf{H}_{\mathrm{IFE}}$ is nearly perpendicular to the
magnetization for this case).

\section{The case $\mathbf{\hat{q}}\parallel\mathbf{H}_{\mathrm{dc}}$}

The Jones matrices corresponding to clockwise and counter-clockwise directions
of loop circulation, are given by $J_{\mathrm{+}}=\sigma_{z}J_{\mathrm{S}%
}\left(  t_{1}\right)  $ and $J_{\mathrm{-}}=\sigma_{z}J_{\mathrm{S}}\left(
t_{2}\right)  \sigma_{z}\sigma_{z}$, respectively, where $J_{\mathrm{S}%
}\left(  t\right)  $ is the FSR Jones matrix at time $t$, and where
$\boldsymbol\sigma=\left(  \sigma_{x},\sigma_{y},\sigma_{z}\right)  $ is the
Pauli matrix vector [see Eq. (\ref{Pauli matrix vector}) and Eqs. (14.106) and
(14.112) of \cite{Buks_QMLN}, and note that the transmission through the loop
gives rise to a mirror reflection of the SOP and that $\sigma_{z}^{2}=1$]. The
term $\chi_{\mathrm{P}}$ is thus given by $\chi_{\mathrm{P}}=\left\langle
p_{\mathrm{i}}\right\vert J_{\mathrm{S}}^{\dag}\left(  t_{1}\right)
J_{\mathrm{S}}\left(  t_{2}\right)  \left\vert p_{\mathrm{i}}\right\rangle $.

Let $\varphi_{\mathrm{S1}}$ and $\varphi_{\mathrm{S2}}$, be the rotation
angles associated with the unitary transformations $J_{\mathrm{S}}\left(
t_{1}\right)  $ and $J_{\mathrm{S}}\left(  t_{2}\right)  $, respectively. For
the case $\mathbf{\hat{q}}\parallel\mathbf{H}_{\mathrm{dc}}$, circular
birefringence (CB) induced by the Faraday effect is the dominant mechanism
giving rise to the change in SOP, and the corresponding Jones matrices
$J_{\mathrm{S}}\left(  t_{1}\right)  $ and $J_{\mathrm{S}}\left(
t_{2}\right)  $ can be calculated using Eq. (\ref{J sphere}) with
$\mathbf{k}_{\mathrm{B}}=\mathbf{k}_{\mathrm{CB}}$ [see Eq. (\ref{k_CB})]. As
is shown in appendix \ref{App_MO}, for the Faraday effect typically
$\left\vert \varphi_{\mathrm{S1}}\right\vert \simeq0.1$ and $\left\vert
\varphi_{\mathrm{S2}}\right\vert \simeq0.1$ for a magnetically saturated FSR
of radius $R_{\mathrm{s}}\simeq100%
%TCIMACRO{\unit{\U{3bc}m}}%
%BeginExpansion
\operatorname{\mu m}%
%EndExpansion
$. Hence, during the time interval $\left[  t_{1},t_{2}\right]  $, the
intermediate value of $\operatorname{Re}\left(  \chi_{\mathrm{P}}\right)  $ is
expected to be significantly smaller than unity.

The final (i.e. after time $t_{2}$) value of $\operatorname{Re}\left(
\chi_{\mathrm{P}}\right)  $ depends on the rotation angle $\varphi
_{\mathrm{S}}$ associated with the unitary transformations $J_{\mathrm{S}%
}^{\dag}\left(  t_{1}\right)  J_{\mathrm{S}}\left(  t_{2}\right)  $. The Jones
matrix $J_{\mathrm{S}}$ given by Eq. (\ref{J sphere}) of appendix \ref{App_MO}
is expressed as a function of the FSR SOM. For the case where FSR excitation
during the time interval $\left(  t_{1},t_{2}\right)  $ is on the order of a
single magnon, one has $\left\vert \varphi_{\mathrm{S}}\right\vert
\simeq\left(  l_{\mathrm{e}}/l_{\mathrm{P}}\right)  \theta_{\mathrm{m0}}$,
where $\theta_{\mathrm{m0}}$ is the magnetization rotation angle corresponding
to a single magnon excitation. As is shown in appendix \ref{App_MO}, typically
$l_{\mathrm{e}}/l_{\mathrm{P}}\simeq10^{-1}$. From the Stoner--Wohlfarth
energy $E_{\mathrm{M}}$ given by Eqs. (\ref{theta_MZ}) and
(\ref{Stoner- Wohlfarth}) one finds that typically $\theta_{\mathrm{m0}}%
\simeq10^{-9}$ (for the transition from the ground state to a single magnon
excitation state). Hence the approximation $\chi_{\mathrm{P}}=1$ (i.e.
$\varphi_{\mathrm{S}}=0$) can be safely employed in the calculation of $\eta$,
provided that the the number of excited magnons is sufficiently small. The
unique configuration of the proposed interferometer allows a finite value of
$\operatorname{Re}\left(  \chi_{\mathrm{P}}\right)  $ very close to unity, in
spite the fact that the intermediate value of $\operatorname{Re}\left(
\chi_{\mathrm{P}}\right)  $ can be significantly smaller than unity.

\section{The case $\mathbf{\hat{q}}\perp\mathbf{H}_{\mathrm{dc}}$}

For simplicity, consider first the case where the FSR is prepared in its
ground state before the optical pulse is applied (i.e. initially the angle
$\theta_{\mathrm{m}}$ between the magnetization and the externally applied
static magnetic field $\mathbf{H}_{\mathrm{dc}}$ vanishes). Let $\theta
_{\mathrm{IFE}}$ be the value of $\theta_{\mathrm{m}}$ immediately after the
interaction with a pulse carrying a single optical photon. The intermediate
value of $\operatorname{Re}\left(  \chi_{\mathrm{M}}\right)  $ during the time
interval $\left[  t_{1},t_{2}\right]  $ is expected to be significantly
smaller than unity provided that $\left\vert \theta_{\mathrm{IFE}}\right\vert
\gtrsim\left\vert \theta_{\mathrm{m0}}\right\vert $ (recall that
$\theta_{\mathrm{m0}}$ is the magnetization rotation angle corresponding to a
single magnon excitation). This condition can be satisfied when angular
momentum conversion between photons and magnons is sufficiently efficient
\cite{Woodford_212412}. On the other hand, as is shown below, the final (i.e.
after time $t_{2}$) value of $\operatorname{Re}\left(  \chi_{\mathrm{M}%
}\right)  $ can be made very close to unity. Note that the semiclassical model
that is presented in appendix \ref{App_MO} allows expressing $\left\vert
\theta_{\mathrm{m0}}\right\vert $ as a function of the magnetization tilt
angle $\theta_{\mathrm{m}}$ and the constant $\theta_{\mathrm{mz}}$ given by
Eq. (\ref{theta_MZ}) [see Eqs. (\ref{Stoner- Wohlfarth}) and (\ref{theta_IFE})].

The level of entanglement associated with the state $\left\vert \psi
_{\mathrm{f}}\right\rangle $ (\ref{|psi_f>}) can be characterized by the
purity $\varrho_{\mathrm{i}}=\operatorname{Tr}\rho_{\mathrm{I}}^{2}%
=\operatorname{Tr}\rho_{\mathrm{M}}^{2}$ of the reduced density matrices
$\rho_{\mathrm{I}}$ and $\rho_{\mathrm{M}}$ of the optical and FSR subsystems,
respectively, which can be extracted from the Schmidt decomposition of
$\left\vert \psi_{\mathrm{i}}\right\rangle $ \cite{Ekert_415}. In the absence
of entanglement $\varrho_{\mathrm{i}}=1$, whereas for a maximized entanglement
$\varrho_{\mathrm{i}}=1/2$. Consider the case of weak excitation, for which
the SOM angle $\theta_{\mathrm{m}}$ is small. For this case, the Bosonization
Holstein-Primakoff transformation \cite{Holstein_1098} can be employed, in
order to allow the description of the state of the transverse magnetization in
terms of a quantum state vector in the Hilbert space of a one-dimensional
harmonic oscillator (i.e. a Boson). Such a description greatly simplifies the
calculation of the purity $\varrho_{\mathrm{i}}$.

Consider the case where the SOP of the partial pulse hitting the FSR at time
$t_{1}$ is adjusted to be circular left-hand $\left\vert L\right\rangle $ SOP.
For that case the partial pulse hitting the FSR at the later time $t_{2}%
>t_{1}$ is expected to have circular right-hand $\left\vert R\right\rangle $
SOP (the loop gives rise to a mirror reflection of the SOP). The precession of
the SOM with angular frequency $\omega_{\mathrm{m}}$ during the time interval
$\left(  t_{1},t_{2}\right)  $ is described by the unitary time evolution
operator $u\left(  t_{2}-t_{1}\right)  $, where $u\left(  t\right)
=\exp\left(  -i\omega_{\mathrm{m}}ta_{\mathrm{m}}^{\dag}a_{\mathrm{m}}\right)
$, and where $a_{\mathrm{m}}$ is a magnon annihilation operator. The change in
the SOM induced by the IFE due to the partial pulse hitting the FSR at time
$t_{1}$ ($t_{2}$) is described by a displacement operator $D\left(
\alpha_{\mathrm{i}}\right)  $\ ($D\left(  -\alpha_{\mathrm{i}}\right)  $),
where the coherent state complex parameter $\alpha_{\mathrm{i}}$ has length
given by $\left\vert \alpha_{\mathrm{i}}\right\vert =\theta_{\mathrm{IFE}%
}/\theta_{\mathrm{m0}}$. It is assumed that $\omega_{\mathrm{m}}t_{\mathrm{p}%
}\ll1$, where $t_{\mathrm{p}}$ is the pulse time duration.

When the initial SOM is assumed to be a coherent state $\left\vert
\alpha\right\rangle $ with a complex parameter $\alpha$, the final SOM
corresponding to circulating the FOLM in the clockwise (counter clockwise)
direction is a coherent state $\left\vert m_{+}\right\rangle _{\mathrm{M}%
}=\left\vert \alpha_{+}\right\rangle $ ($\left\vert m_{-}\right\rangle
_{\mathrm{M}}=\left\vert \alpha_{-}\right\rangle $) with complex parameter
$\alpha_{+}=\left(  \alpha+\alpha_{\mathrm{i}}\right)  e^{-i\omega
_{\mathrm{m}}\left(  t_{2}-t_{1}\right)  }$ ($\alpha_{-}=\alpha e^{-i\omega
_{\mathrm{m}}\left(  t_{2}-t_{1}\right)  }-\alpha_{\mathrm{i}}$) [see Eq.
(5.53) of \cite{Buks_QMLN}]. The state vector $\left\vert \psi_{\mathrm{f}%
}\right\rangle $ can be expressed as $\left\vert \psi_{\mathrm{f}%
}\right\rangle =v_{1}\left\vert a_{1}\longleftarrow\right\rangle _{\mathrm{I}%
}\otimes\left\vert m_{1}\right\rangle _{\mathrm{M}}+v_{2}\left\vert
a_{2}\longleftarrow\right\rangle _{\mathrm{I}}\otimes\left\vert m_{2}%
\right\rangle _{\mathrm{M}}$ , where $v_{1}=it^{2}\sqrt{\upsilon\nu_{+}}$,
$\left\vert m_{1}\right\rangle _{\mathrm{M}}=\left(  \left\vert \alpha
_{+}\right\rangle +\left\vert \alpha_{-}\right\rangle \right)  /\sqrt{\nu_{+}%
}$, $v_{2}=t^{2}\sqrt{\left(  1-\upsilon\right)  ^{2}+\upsilon\nu_{-}}$,
$\left\vert m_{2}\right\rangle _{\mathrm{M}}=\left(  \left\vert \alpha
_{+}\right\rangle -\upsilon\left\vert \alpha_{-}\right\rangle \right)
/\sqrt{\left(  1-\upsilon\right)  ^{2}+\upsilon\nu_{-}}$, $\mu=\left\langle
\alpha_{+}\right.  \left\vert \alpha_{-}\right\rangle =\mu^{\prime}%
+i\mu^{\prime\prime}$, with both $\mu^{\prime}$ and $\mu^{\prime\prime}%
$\ being real, $\upsilon=\left\vert r/t\right\vert ^{2}$ and $\nu_{\pm
}=2\left(  1\pm\mu^{\prime}\right)  $ [see Eq. (\ref{|psi_f>})]. Note that
both $\left\vert m_{1}\right\rangle _{\mathrm{M}}$ and $\left\vert
m_{2}\right\rangle _{\mathrm{M}}$\ are normalized. The purity $\varrho
_{\mathrm{i}}$ associated with the state $\left\vert \psi_{\mathrm{f}%
}\right\rangle $ is given by $\varrho_{\mathrm{i}}=1-2\left\vert v_{1}%
v_{2}\right\vert ^{2}\left(  1-\left\vert _{\mathrm{M}}\left\langle
m_{1}\right.  \left\vert m_{2}\right\rangle _{\mathrm{M}}\right\vert
^{2}\right)  $ [see Eq. (8.681) of \cite{Buks_QMLN}]. For a 3 dB OC, i.e. for
$\upsilon=\left\vert r/t\right\vert ^{2}=1$, this becomes $\varrho
_{\mathrm{i}}=\left(  1+\exp\left(  -\left\vert \alpha_{+}-\alpha
_{-}\right\vert ^{2}\right)  \right)  /2$ [see Eq. (5.243) of \cite{Buks_QMLN}%
], or (note that $\varrho_{\mathrm{i}}$ is independent on $\alpha$)%
\begin{equation}
\varrho_{\mathrm{i}}=\frac{1+\exp\left(  -4\left\vert \alpha_{\mathrm{i}%
}\right\vert ^{2}\cos^{2}\frac{\omega_{\mathrm{m}}\left(  t_{2}-t_{1}\right)
}{2}\right)  }{2}\;. \label{rho_i}%
\end{equation}

The time interval $t_{2}-t_{1}$ can be set by adjusting the length of the
fiber loop connecting ports b1 and b2 of the OC. A delay time of a single FSR
period $\omega_{\mathrm{m}}/\left(  2\pi\right)  $ is obtained with fiber
having length $L_{\mathrm{F}}$ given by $L_{\mathrm{F}}=cn_{\mathrm{F}}%
^{-1}\left(  \omega_{\mathrm{m}}/\left(  2\pi\right)  \right)  ^{-1}=68%
%TCIMACRO{\unit{mm}}%
%BeginExpansion
\operatorname{mm}%
%EndExpansion
\left(  n_{\mathrm{F}}/1.47\right)  ^{-1}\left(  \left(  \omega_{\mathrm{m}%
}/\left(  2\pi\right)  \right)  /\left(  3%
%TCIMACRO{\unit{GHz}}%
%BeginExpansion
\operatorname{GHz}%
%EndExpansion
\right)  \right)  ^{-1}$, where $n_{\mathrm{F}}$ is the fiber's effective
refractive index. When the ratio $\left(  t_{2}-t_{1}\right)  /\left(
2\pi/\omega_{\mathrm{m}}\right)  $ is much smaller than the FSR quality factor
the effect of magnon damping can be disregarded.

During the time interval $\left(  t_{1},t_{2}\right)  $ the entanglement is
nearly maximized provided that $e^{-\left\vert \alpha_{\mathrm{i}}\right\vert
^{2}}\ll1$. For a symmetric OC (i.e. for $\left\vert r/t\right\vert =1$), a
full collapse accruing during this time interval results in a transmission
probability $p_{\mathrm{T}}\simeq1/2$, whereas unitary evolution yields
$p_{\mathrm{T}}\simeq0$. Consider the case where the condition $\cos\left(
\omega_{\mathrm{m}}\left(  t_{2}-t_{1}\right)  /2\right)  =0$ is satisfied.
Note that for this case $u\left(  t_{2}-t_{1}\right)  \left\vert
\alpha\right\rangle =\left\vert -\alpha\right\rangle $, hence the partial
pulse hitting the FSR at time $t_{2}$ undoes the earlier change that has
occurred at time $t_{1}$ (recall that the fiber loop gives rise to a mirror
transformation $\left\vert L\right\rangle \rightarrow\left\vert R\right\rangle
$ in the SOP), and consequently entanglement is eliminated, and the final
state of the system $\left\vert \psi_{\mathrm{f}}\right\rangle $ after time
$t_{2}$\ becomes a product state, i.e. $\operatorname{Re}\left(
\chi_{\mathrm{M}}\right)  =1$

In the analysis above the Sagnac effect has been disregarded. In general, this
effect, which gives rise to a relative phase shift between the clockwise and
counter-clockwise partial pulses, can also contribute to the suppression of
the destructive interference at the outgoing OC port a2. The Sagnac effect can
be eliminated by placing the fiber loop in a plane parallel to the earth
rotation axis.

\section{Summary}

Devices similar to the one discussed here, which are based on ferrimagnetic MO
coupling
\cite{Almpanis_184406,Hisatomi_207401,Pantazopoulos_104425,Sharma_094412,Hisatomi_174427}%
, are currently being developed worldwide
\cite{Lachance_070101,Wolski_2005_09250,Zhu_2005_06429}, mainly for the
purpose of optically interfacing superconducting quantum circuits. Ultrafast
(sub $%
%TCIMACRO{\unit{ps}}%
%BeginExpansion
\operatorname{ps}%
%EndExpansion
$ time scales) laser control of the SOM \cite{Kimel_275} can be employed for
the preparation and manipulation of non-classical states of a FSR.

The device we propose here is designed to allow studying the quantum to
classical transition associated with the interaction between an optical pulse
and a FSR containing $\sim10^{17}$ spins. The measured transmission
probability $p_{\mathrm{T}}$ provides a very sensitive probe for non-unitarity
in the system's time evolution. Unitary evolution\ yields $p_{\mathrm{T}%
}\simeq0$, whereas a full collapse occurring during the time interval $\left(
t_{1},t_{2}\right)  $ results in $p_{\mathrm{T}}\simeq1/2$. The proposed
experimental setup allows the generation of an entangled state during the time
interval $\left(  t_{1},t_{2}\right)  $. The level of entanglement after time
$t_{2}$ can be controlled by adjusting the time duration $t_{1}-t_{2}$ (which
can be made much shorter than all time scales characterizing environmental
decoherence). Systematic measurements of the transmission probability
$p_{\mathrm{T}}$ with varying parameters may provide an important insight on
the non-unitary nature of a quantum measurement.

\section{Acknowledgments}

This work was supported by the Israeli science foundation, the Israeli
ministry of science, and by the Technion security research foundation.

\appendix

\section{Magneto-optics}

\label{App_MO}

In this appendix the MO Faraday, Voigt and inverse Faraday effects are briefly reviewed.

\subsection{Macroscopic Maxwell's equations}

In the absent of current sources, the macroscopic Maxwell's equations in
Fourier space are given by%
\begin{align}
i\mathbf{q}\times\mathbf{H}_{\mathrm{T}}\left(  \mathbf{q},\omega\right)   &
=-\frac{i\omega}{c}\mathbf{D}\left(  \mathbf{q},\omega\right)  \;,
\label{MMF1}\\
\mathbf{q}\times\mathbf{E}_{\mathrm{T}}\left(  \mathbf{q},\omega\right)   &
=\frac{\omega}{c}\mathbf{B}\left(  \mathbf{q},\omega\right)  \;,\label{MMF2}\\
i\mathbf{q}\cdot\mathbf{D}_{\mathrm{L}}\left(  \mathbf{q},\omega\right)   &
=4\pi\rho_{\mathrm{ext}}\left(  \mathbf{q},\omega\right)  \;,\label{MMF3}\\
\mathbf{q}\cdot\mathbf{B}_{\mathrm{L}}\left(  \mathbf{q},\omega\right)   &
=0\;, \label{MMF4}%
\end{align}
where $\mathbf{H}$ is the magnetic field, $\mathbf{E}$ is the electric field,
$\mathbf{B}$ is the magnetic induction, $\mathbf{D}$ is the electric
displacement, $\rho_{\mathrm{ext}}$ is the charge density, $c$\ is the speed
of light, $\mathbf{q}$ is the Fourier wave vector, and $\omega$ is the Fourier
angular frequency. All vector fields $\mathbf{F}\in\left\{  \mathbf{H}%
,\mathbf{E},\mathbf{B},\mathbf{D}\right\}  $ are decomposed into longitudinal
and transverse parts with respect to the wave vector $\mathbf{q}$ according to
$\mathbf{F}=\mathbf{F}_{\mathrm{L}}+\mathbf{F}_{\mathrm{T}}$. where the
longitudinal part is given by $\mathbf{F}_{\mathrm{L}}=\left(  \mathbf{\hat
{q}}\cdot\mathbf{F}\right)  \mathbf{\hat{q}}$, the transverse one is given by
$\mathbf{F}_{\mathrm{T}}=\left(  \mathbf{\hat{q}}\times\mathbf{F}\right)
\times\mathbf{\hat{q}}$, and where $\mathbf{\hat{q}}=\mathbf{q/}\left\vert
\mathbf{q}\right\vert $ is a unit vector in the direction of $\mathbf{q}$. For
an isotropic and linear medium the following relations hold $\mathbf{D}%
=\epsilon_{\mathrm{m}}\mathbf{E}$, where $\epsilon_{\mathrm{m}}$ is the
permittivity tensor, and $\mathbf{B}=\mu_{\mathrm{m}}\mathbf{H}$, where
$\mu_{\mathrm{m}}$ is the permeability tensor. In the optical band to a good
approximation $\mu_{\mathrm{m}}$ is the identity tensor.

By applying $\mathbf{q}\times$ to Eq. (\ref{MMF2}) from the left, and
employing Eq. (\ref{MMF1}) one obtains $\mathbf{q}\times\left(  \mathbf{q}%
\times\mathbf{E}_{\mathrm{T}}\right)  =-\epsilon\left(  \omega/c\right)
^{2}\mathbf{E}_{\mathrm{T}}$ \cite{Freiser_152,Boardman_197,Boardman_388}, or
in a matrix form [note that for general vectors $\mathbf{u}$ and $\mathbf{v}$
the following holds $\mathbf{u}\times\left(  \mathbf{u}\times\mathbf{v}%
\right)  =\left(  \mathbf{uu}^{\mathrm{T}}-\mathbf{u}\cdot\mathbf{u}\right)
\mathbf{v}$]%
\begin{equation}
\left(  M_{\epsilon}+1-\frac{n^{2}}{n_{0}^{2}}\right)  \mathbf{E}_{\mathrm{T}%
}=0\;, \label{()*E=0}%
\end{equation}
where the $3\times3$ matrix $M_{\epsilon}$ is given by%
\begin{equation}
M_{\epsilon}=\frac{\epsilon_{\mathrm{m}}}{n_{0}^{2}}+\frac{\mathbf{qq}%
^{\mathrm{T}}}{n_{0}^{2}q_{0}^{2}}-1=\frac{\epsilon_{\mathrm{m}}%
+n^{2}P_{\mathbf{\hat{q}}}}{n_{0}^{2}}-1\;, \label{M_epsilon}%
\end{equation}
$\mathbf{q}=q\mathbf{\hat{q}}$, $\mathbf{\hat{q}}=\left(  \sin\theta\cos
\phi,\sin\theta\sin\phi,\cos\theta\right)  $, $q_{0}=\omega/c$, $n_{0}$ is the
medium refractive index, $n=q/q_{0}$, and where $P_{\mathbf{\hat{u}}%
}=\mathbf{\hat{u}\hat{u}}^{\mathrm{T}}$ is a projection matrix associated with
a given unit vector $\mathbf{\hat{u}}$ (the $3\times3$ identity matrix is
denoted by $1$). Note that $n^{2}/n_{0}^{2}-1\simeq2\left(  n-n_{0}\right)
/n_{0}$ provided that $\left\vert n-n_{0}\right\vert \ll n_{0}$.

For a ferromagnet or a ferrimagnet medium, it is assumed that the elements
$\epsilon_{ij}$ are functions of the magnetization vector $\mathbf{M}$. The
Onsager's time-reversal symmetry relation reads $\epsilon_{ij}\left(
\mathbf{M}\right)  =\epsilon_{ji}\left(  -\mathbf{M}\right)  $. Moreover, it
is expected that $\epsilon_{ij}\left(  \mathbf{M}=0\right)  =0$ for $i\neq j$.
The static magnetic field $\mathbf{H}_{\mathrm{dc}}$ is assumed to be parallel
to the $\mathbf{\hat{z}}$ direction. For the case where $\mathbf{M}$ is
parallel to $\mathbf{H}_{\mathrm{dc}}$ (i.e. parallel to $\mathbf{\hat{z}}$)
the tensor $\epsilon_{\mathrm{m}}$ is assumed to have the form
\cite{Boardman_197,Boardman_388}%
\begin{equation}
\frac{\epsilon_{\mathrm{m}}}{n_{0}^{2}}=1+iQM_{\mathrm{C}}\;, \label{ep_m}%
\end{equation}
where the matrix $M_{\mathrm{C}}$ is given by%
\begin{equation}
M_{\mathrm{C}}=\left(
\begin{array}
[c]{ccc}%
0 & -1 & 0\\
1 & 0 & 0\\
0 & 0 & 0
\end{array}
\right)  \;.
\end{equation}

The value of $Q$ corresponding to saturated magnetization is denoted by
$Q_{\mathrm{s}}$. For YIG $Q_{\mathrm{s}}\simeq10^{-4}$ for (free space)
wavelength $\lambda_{0}\simeq1550%
%TCIMACRO{\unit{nm}}%
%BeginExpansion
\operatorname{nm}%
%EndExpansion
$ in the telecom band \cite{Wood_1038}. The corresponding polarization beat
length $l_{\mathrm{P}}$ is given by $l_{\mathrm{P}}=\lambda_{0}/\left(
n_{0}Q_{\mathrm{s}}\right)  \simeq7.\,0%
%TCIMACRO{\unit{mm}}%
%BeginExpansion
\operatorname{mm}%
%EndExpansion
$, where $n_{0}=2.19$ is the refractive index of YIG in the telecom band. In
this band $l_{\mathrm{P}}/l_{\mathrm{A}}\simeq0.014\,$, where $l_{\mathrm{A}%
}^{-1}=\left(  0.5%
%TCIMACRO{\unit{m}}%
%BeginExpansion
\operatorname{m}%
%EndExpansion
\right)  ^{-1}$ is the YIG absorption coefficient
\cite{Zhang_591,Onbasli_1,Donati_372,Jooss_651}.

To analyze the change in the SOP induced by MO coupling, a rotation
transformation is applied to a coordinate system having a $z$ axis parallel to
the propagation direction ($\mathbf{\hat{q}}$ in the non-rotated frame). Let
$M_{\epsilon}^{\prime}$ be the transformed matrix that represents the matrix
$M_{\epsilon}$ in that coordinate system. For a given unit vector
$\mathbf{\hat{u}}$, the rotation matrix $R_{\mathbf{\hat{u}}}$ is defined by
the relation $R_{\mathbf{\hat{u}}}\mathbf{\hat{u}}=\mathbf{\hat{z}}$. The unit
vector parallel to the magnetization $\mathbf{M}$ is denoted by $\mathbf{\hat
{m}}=\left(  \sin\theta_{\mathrm{m}}\cos\phi_{\mathrm{m}},\sin\theta
_{\mathrm{m}}\sin\phi_{\mathrm{m}},\cos\theta_{\mathrm{m}}\right)  $. The
transformed matrix $M_{\epsilon}^{\prime}$ is given by%
\begin{equation}
M_{\epsilon}^{\prime}=\frac{R_{\mathbf{\hat{q}}}R_{\mathbf{\hat{m}}}%
^{-1}\epsilon_{\mathrm{m}}R_{\mathbf{\hat{m}}}R_{\mathbf{\hat{q}}}^{-1}%
+n^{2}P_{\mathbf{\hat{z}}}}{n_{0}^{2}}-1\;. \label{M'_epsilon}%
\end{equation}
Note that Eq. (\ref{M'_epsilon}) implies that (note that $R_{\mathbf{\hat{q}}%
}^{-1}\mathbf{\hat{z}}=\mathbf{\hat{q}}$ and $R_{\mathbf{\hat{u}}}%
^{-1}=R_{\mathbf{\hat{u}}}^{\mathrm{T}}$)%
\begin{equation}
R_{\mathbf{\hat{q}}}^{-1}M_{\epsilon}^{\prime}R_{\mathbf{\hat{q}}}%
=\frac{R_{\mathbf{\hat{m}}}^{-1}\epsilon_{\mathrm{m}}R_{\mathbf{\hat{m}}%
}+n^{2}P_{\mathbf{\hat{q}}}}{n_{0}^{2}}-1\;,
\end{equation}
and%
\begin{equation}
R_{\mathbf{\hat{m}}}R_{\mathbf{\hat{q}}}^{-1}M_{\epsilon}^{\prime
}R_{\mathbf{\hat{q}}}R_{\mathbf{\hat{m}}}^{-1}=\frac{\epsilon_{\mathrm{m}%
}+n^{2}R_{\mathbf{\hat{m}}}P_{\mathbf{\hat{q}}}R_{\mathbf{\hat{m}}}^{-1}%
}{n_{0}^{2}}-1\;.
\end{equation}
Note also that [see Eq. (6.235) of \cite{Buks_QMLN}]%
\begin{equation}
\frac{R_{\mathbf{\hat{m}}}^{-1}\left(  \frac{\epsilon_{\mathrm{m}}}{n_{0}^{2}%
}-1\right)  R_{\mathbf{\hat{m}}}}{iQ_{\mathrm{s}}}=R_{\mathbf{\hat{m}}}%
^{-1}M_{\mathrm{C}}R_{\mathbf{\hat{m}}}=C_{\mathbf{\hat{m}}}\;, \label{M_C R}%
\end{equation}
where the matrix $C_{\mathbf{\hat{u}}}$, which is defined by%
\begin{equation}
C_{\mathbf{\hat{u}}}=\left(
\begin{array}
[c]{ccc}%
0 & -\mathbf{\hat{u}}\cdot\mathbf{\hat{z}} & \mathbf{\hat{u}}\cdot
\mathbf{\hat{y}}\\
\mathbf{\hat{u}}\cdot\mathbf{\hat{z}} & 0 & -\mathbf{\hat{u}}\cdot
\mathbf{\hat{x}}\\
-\mathbf{\hat{u}}\cdot\mathbf{\hat{y}} & \mathbf{\hat{u}}\cdot\mathbf{\hat{x}}
& 0
\end{array}
\right)  \;, \label{C_m}%
\end{equation}
is the cross-product matrix corresponding to a given unit vector
$\mathbf{\hat{u}}$, and for an arbitrary 3-dimensional vector $\mathbf{v}$ the
following holds $\mathbf{\hat{u}\times v=}C_{\mathbf{\hat{u}}}\mathbf{v}$ [see
Eq. (6.243) of \cite{Buks_QMLN}]. The following holds%
\begin{equation}
C_{\mathbf{\hat{m}}}=M_{\mathrm{C}}+M_{\perp}+O\left(  \theta_{\mathrm{m}}%
^{2}\right)  \;,
\end{equation}
where the matrix $M_{\perp}$\ is given by%
\begin{equation}
M_{\perp}=\theta_{\mathrm{m}}\left(
\begin{array}
[c]{ccc}%
0 & 0 & \sin\phi_{\mathrm{m}}\\
0 & 0 & -\cos\phi_{\mathrm{m}}\\
-\sin\phi_{\mathrm{m}} & \cos\phi_{\mathrm{m}} & 0
\end{array}
\right)  \;,
\end{equation}
hence to first order in $\theta_{\mathrm{m}}$ one has [see Eq.
(\ref{M'_epsilon}), and note that the approximation $\left(  n^{2}/n_{0}%
^{2}\right)  P_{\mathbf{\hat{z}}}\simeq P_{\mathbf{\hat{z}}}$ is being
employed]%
\begin{equation}
M_{\epsilon}^{\prime}=iQ_{\mathrm{s}}R_{\mathbf{\hat{q}}}\left(
M_{\mathrm{C}}+M_{\perp}\right)  R_{\mathbf{\hat{q}}}^{-1}+P_{\mathbf{\hat{z}%
}}\;,
\end{equation}
or [compare with Eq. (\ref{M_C R})]%
\begin{equation}
M_{\epsilon}^{\prime}=\left(
\begin{array}
[c]{ccc}%
0 & -iQ_{z} & -iQ_{y}\\
iQ_{z} & 0 & iQ_{x}\\
iQ_{y} & -iQ_{x} & 1
\end{array}
\right)  +iQ_{\mathrm{s}}R_{\mathbf{\hat{q}}}M_{\perp}R_{\mathbf{\hat{q}}%
}^{-1}\;,
\end{equation}
where $\left(  Q_{x},Q_{y},Q_{z}\right)  =Q_{\mathrm{s}}\mathbf{\hat{q}}$.

An effective $2\times2$ matrix $M_{\mathrm{T}}$ corresponding to the
transverse components of the electric field (spanned by the first two vectors)
is evaluated below using Eq. (4.87) of \cite{Buks_QMLN}. \ When terms of
orders $\theta_{\mathrm{m}}Q_{\mathrm{s}}^{2}$ are disregarded (it is assumed
that $\left\vert \theta_{\mathrm{m}}\right\vert \ll1$ and $Q_{\mathrm{s}}\ll
1$), one finds using the relation%
\begin{equation}
\frac{\left(
\begin{array}
[c]{ccc}%
1 & 0 & 0\\
0 & 1 & 0\\
0 & 0 & 0
\end{array}
\right)  R_{\mathbf{\hat{q}}}M_{\perp}R_{\mathbf{\hat{q}}}^{-1}\left(
\begin{array}
[c]{ccc}%
1 & 0 & 0\\
0 & 1 & 0\\
0 & 0 & 0
\end{array}
\right)  }{\theta_{\mathrm{m}}\cos\left(  \phi-\phi_{\mathrm{m}}\right)
\sin\theta}=M_{\mathrm{C}}\;,
\end{equation}
that%
\[
M_{\mathrm{T}}=Q_{\mathrm{s}}\alpha_{\mathrm{CB}}\left(
\begin{array}
[c]{cc}%
0 & -i\\
i & 0
\end{array}
\right)  +\left(
\begin{array}
[c]{cc}%
-Q_{y}^{2} & Q_{x}Q_{y}\\
Q_{x}Q_{y} & -Q_{x}^{2}%
\end{array}
\right)  \;,
\]
where $\alpha_{\mathrm{CB}}$ is given by [recall that $\cos\left(  \phi
-\phi_{\mathrm{m}}\right)  =\cos\phi\cos\phi_{\mathrm{m}}+\sin\phi\sin
\phi_{\mathrm{m}}$]%
\begin{equation}
\alpha_{\mathrm{CB}}=\frac{Q_{z}}{Q_{\mathrm{s}}}+\theta_{\mathrm{m}}%
\cos\left(  \phi-\phi_{\mathrm{m}}\right)  \sin\theta=\mathbf{\hat{q}}%
\cdot\mathbf{\hat{m}}+O\left(  \theta_{\mathrm{m}}^{2}\right)  \;,
\end{equation}
or%
\begin{equation}
M_{\mathrm{T}}=k_{0}\sigma_{0}+\mathbf{k}_{\mathrm{B}}\cdot\mathbf{\sigma}\;,
\label{M_T sigma}%
\end{equation}
where $k_{0}=-\left(  Q_{x}^{2}+Q_{y}^{2}\right)  /2$, $\sigma_{0}$ is the
$2\times2$ identity matrix, the Pauli matrix vector $\boldsymbol\sigma=\left(
\sigma_{x},\sigma_{y},\sigma_{z}\right)  $ is given by%
\begin{equation}
\sigma_{x}=\left(
\begin{array}
[c]{cc}%
0 & 1\\
1 & 0
\end{array}
\right)  ,\;\sigma_{y}=\left(
\begin{array}
[c]{cc}%
0 & -i\\
i & 0
\end{array}
\right)  ,\;\sigma_{z}=\left(
\begin{array}
[c]{cc}%
1 & 0\\
0 & -1
\end{array}
\right)  \;, \label{Pauli matrix vector}%
\end{equation}
the birefringence vector $\mathbf{k}_{\mathrm{B}}$ is expressed as
$\mathbf{k}_{\mathrm{B}}=\mathbf{k}_{\mathrm{CB}}+\mathbf{k}_{\mathrm{LB}}$,
with (to first order in $\theta_{\mathrm{m}}$)%
\begin{equation}
\mathbf{k}_{\mathrm{CB}}=Q_{\mathrm{s}}\left(  0,\mathbf{\hat{q}}%
\cdot\mathbf{\hat{m}},0\right)  \;, \label{k_CB}%
\end{equation}
and%
\begin{equation}
\mathbf{k}_{\mathrm{LB}}=Q_{\mathrm{s}}^{2}\left(  S\left(  -\frac{\pi}%
{4}\right)  ,0,S\left(  \frac{\pi}{4}\right)  \right)  \;, \label{k_LB}%
\end{equation}
where the squeezing transformation $S\left(  \varrho\right)  $ is given by%
\begin{equation}
S\left(  \varrho\right)  =\frac{e^{i\left(  \varrho-\frac{\pi}{4}\right)
}\mathcal{Q}^{2}+e^{-i\left(  \varrho-\frac{\pi}{4}\right)  }\mathcal{Q}%
^{\ast2}}{4}\;,
\end{equation}
and where $\mathcal{Q}=\left(  Q_{x}+iQ_{y}\right)  /Q_{\mathrm{s}}$.

\subsection{Jones matrices}

In general, the transformation between input SOP and output SOP for a given
optical element can be described using a Jones matrix $J$ \cite{Potton_717}.
For the loss-less case the matrix $J$ is unitary, and it can be expressed as
$J=B\left(  \mathbf{\hat{u}},\varphi\right)  $, where%
\begin{equation}
B\left(  \mathbf{\hat{u}},\varphi\right)  \dot{=}\exp\left(  -\frac
{i\mathbf{\sigma}\cdot\mathbf{\hat{u}}\varphi}{2}\right)  =\mathbf{1}\cos
\frac{\varphi}{2}-i\mathbf{\sigma}\cdot\mathbf{\hat{u}}\sin\frac{\varphi}%
{2}\;, \label{B(n,phi)}%
\end{equation}
and where $\mathbf{\hat{u}}$ is a unit vector and $\varphi$ is a rotation
angle. The colinear vertical, horizontal, diagonal and anti-diagonal SOP are
denoted by $\left\vert V\right\rangle $, $\left\vert H\right\rangle $,
$\left\vert D\right\rangle =2^{-1/2}\left(  \left\vert H\right\rangle
+\left\vert V\right\rangle \right)  $ and $\left\vert A\right\rangle
=2^{-1/2}\left(  \left\vert H\right\rangle -\left\vert V\right\rangle \right)
$ , respectively, whereas the circular right-hand and left-hand SOP are
denoted by $\left\vert R\right\rangle =2^{-1/2}\left(  \left\vert
H\right\rangle -i\left\vert V\right\rangle \right)  $ and $\left\vert
L\right\rangle =2^{-1/2}\left(  \left\vert H\right\rangle +i\left\vert
V\right\rangle \right)  $ , respectively. The unit vectors in the Poincar\'{e}
sphere corresponding to the SOP $\left\vert V\right\rangle $, $\left\vert
H\right\rangle $, $\left\vert D\right\rangle $, $\left\vert A\right\rangle $,
$\left\vert R\right\rangle $ and $\left\vert L\right\rangle $, are
$\mathbf{\hat{z}}$, $-\mathbf{\hat{z}}$, $\mathbf{\hat{x}}$, $-\mathbf{\hat
{x}}$, $-\mathbf{\hat{y}}$ and $\mathbf{\hat{y}}$, respectively.

Consider a FSR having radius $R_{\mathrm{s}}$ and saturated magnetization.
When damping is disregarded the sphere's Jones matrix $J_{\mathrm{S}}$ is
given by [see Eqs. (\ref{M_T sigma}) and (\ref{B(n,phi)})]%
\begin{equation}
J_{\mathrm{S}}=B\left(  \frac{\mathbf{k}_{\mathrm{B}}}{\left\vert
\mathbf{k}_{\mathrm{B}}\right\vert },\frac{l_{\mathrm{e}}}{l_{\mathrm{P}}%
}\frac{\left\vert \mathbf{k}_{\mathrm{B}}\right\vert }{Q_{\mathrm{s}}}\right)
\;, \label{J sphere}%
\end{equation}
where $l_{\mathrm{e}}\simeq2R_{\mathrm{s}}$ is the effective optical travel
length inside the sphere. The first order in $Q_{\mathrm{s}}$ component of
$\mathbf{k}_{\mathrm{B}}=\mathbf{k}_{\mathrm{CB}}+\mathbf{k}_{\mathrm{LB}}$ in
the $y$ direction [see Eq. (\ref{k_CB})] gives rise to CB known as the Faraday
effect, whereas the second order in $Q_{\mathrm{s}}$ components in the $xz$
plane give rise to colinear birefringence\ (LB) known as the Voigt
(Cotton-Mouton) effect [see Eq. (\ref{k_LB})]. The eigenvectors corresponding
to CB (LB) have circular (colinear) polarization.

\subsection{Stoner--Wohlfarth energy}

When anisotropy is disregarded, the Stoner--Wohlfarth energy $E_{\mathrm{M}}$
of the FSR is given by $E_{\mathrm{M}}=-\mu_{0}V_{\mathrm{s}}M_{\mathrm{s}%
}H_{\mathrm{dc}}\cos\theta_{\mathrm{m}}$, where $\mu_{0}$ is the free space
permeability, $V_{\mathrm{s}}=4\pi R_{\mathrm{s}}^{3}/3$ is the volume of the
sphere having radius $R_{\mathrm{s}}$, $M_{\mathrm{s}}$ is the saturation
magnetization ($M_{\mathrm{s}}=140%
%TCIMACRO{\unit{kA}}%
%BeginExpansion
\operatorname{kA}%
%EndExpansion
/%
%TCIMACRO{\unit{m}}%
%BeginExpansion
\operatorname{m}%
%EndExpansion
$ for YIG at room temperature), $H_{\mathrm{dc}}$ is the static magnetic
field, which is related to the angular frequency of the Kittel mode
$\omega_{\mathrm{m}}$ by $H_{\mathrm{dc}}=\omega_{\mathrm{m}}/\left(  \mu
_{0}\gamma_{\mathrm{e}}\right)  $ \cite{Fletcher_687,sharma2019cavity}, and
$\theta_{\mathrm{m}}$ is the angle between the magnetization and static
magnetic field vectors \cite{Stancil_Spin}. In terms of the angle
$\theta_{\mathrm{mz}}$, which is given by%
\begin{equation}
\theta_{\mathrm{mz}}=\frac{2\hbar\gamma_{\mathrm{e}}}{V_{\mathrm{s}%
}M_{\mathrm{s}}}=\frac{3.2\times10^{-17}}{\left(  \frac{R_{\mathrm{s}}}{125%
%TCIMACRO{\unit{\U{3bc}m}}%
%BeginExpansion
\operatorname{\mu m}%
%EndExpansion
}\right)  ^{3}\frac{M_{\mathrm{s}}}{140%
%TCIMACRO{\unit{kA}}%
%BeginExpansion
\operatorname{kA}%
%EndExpansion
/%
%TCIMACRO{\unit{m}}%
%BeginExpansion
\operatorname{m}%
%EndExpansion
}}\;, \label{theta_MZ}%
\end{equation}
the energy $E_{\mathrm{M}}$ can be expressed as%
\begin{equation}
E_{\mathrm{M}}=-2\hbar\omega_{\mathrm{m}}\frac{\cos\theta_{\mathrm{m}}}%
{\theta_{\mathrm{mz}}}\;. \label{Stoner- Wohlfarth}%
\end{equation}

\subsection{IFE effective magnetic field}

Consider the case where the second order in $Q_{\mathrm{s}}$ LB induced by the
Voigt effect can be disregarded. For this case, for which $\mathbf{k}%
_{\mathrm{B}}$ becomes parallel to the $\mathbf{\hat{y}}$ direction in the
Poincar\'{e} space, it is convenient to express the transverse electric field
in the basis of circular SOP $\mathbf{E}_{\mathrm{T}}^{\prime}=E_{+}%
\mathbf{\hat{u}}_{+}+E_{-}\mathbf{\hat{u}}_{-}$, where $\mathbf{\hat{u}}_{\pm
}=\left(  e^{\mp i\pi/4}/\sqrt{2},e^{\pm i\pi/4}/\sqrt{2}\right)
^{\mathrm{T}}$ (note that $\sigma_{y}\mathbf{\hat{u}}_{\pm}=\pm\mathbf{\hat
{u}}_{\pm}$). For this case the electric energy density $u_{\mathrm{E}%
}=\left(  \epsilon_{0}/2\right)  \left(  \mathbf{E}_{\mathrm{T}}^{\prime\dag
}\epsilon_{\mathrm{m}}^{\prime}\mathbf{E}_{\mathrm{T}}^{\prime}\right)  $ can
be expressed as [see Eqs. (\ref{M'_epsilon}), (\ref{M_T sigma}) and
(\ref{k_CB})]%
\begin{equation}
u_{\mathrm{E}}=\epsilon_{0}\frac{n_{+}^{2}\left\vert E_{+}\right\vert
^{2}+n_{-}^{2}\left\vert E_{-}\right\vert ^{2}}{2}\;,\label{U_E}%
\end{equation}
where $\left\vert E_{+}\right\vert ^{2}$ ($\left\vert E_{-}\right\vert ^{2}$)
is proportional to the intensity of right-hand $\left\vert R\right\rangle $
(left-hand $\left\vert L\right\rangle $) circular SOP, $n_{\pm}=n_{0}\left(
1\pm\left\vert \mathbf{k}_{\mathrm{CB}}\right\vert \right)  ^{1/2}$, and
$\left\vert \mathbf{k}_{\mathrm{CB}}\right\vert =Q_{\mathrm{s}}\left\vert
\mathbf{\hat{q}}\cdot\mathbf{\hat{m}}\right\vert $. Alternatively,
$u_{\mathrm{E}}\ $can be expressed as $u_{\mathrm{E}}=u_{\mathrm{E0}%
}+u_{\mathrm{E1}}$, where $u_{\mathrm{E0}}=\left(  \epsilon_{0}n_{0}%
^{2}/2\right)  \left(  \left\vert E_{+}\right\vert ^{2}+\left\vert
E_{-}\right\vert ^{2}\right)  $ and $u_{\mathrm{E1}}=\left(  \epsilon_{0}%
n_{0}^{2}\left\vert \mathbf{k}_{\mathrm{CB}}\right\vert /2\right)  \left(
\left\vert E_{+}\right\vert ^{2}-\left\vert E_{-}\right\vert ^{2}\right)  $.
When the energy density is uniformly distributed inside the FSR, the energy
$U_{\mathrm{T}}=V_{\mathrm{s}}u_{\mathrm{E1}}$ is given by $U_{\mathrm{T}%
}=\hbar\omega_{\mathrm{e}}\left\vert \mathbf{k}_{\mathrm{CB}}\right\vert
=\hbar\omega_{\mathrm{e}}Q_{\mathrm{s}}\left(  \mathbf{\hat{q}}\cdot
\mathbf{\hat{m}}\right)  $ [see Eq. (\ref{k_CB})] where%
\begin{equation}
\omega_{\mathrm{e}}=\frac{\epsilon_{0}n_{0}^{2}V_{\mathrm{s}}\left(
\left\vert E_{+}\right\vert ^{2}-\left\vert E_{-}\right\vert ^{2}\right)
}{2\hbar}\;,
\end{equation}
or%
\begin{equation}
U_{\mathrm{T}}=\frac{\mu_{0}}{2}\mathbf{H}_{\mathrm{IFE}}\cdot\mathbf{M}\;,
\end{equation}
where the IFE effective magnetic field $\mathbf{H}_{\mathrm{IFE}}$ is given by%
\begin{equation}
\mathbf{H}_{\mathrm{IFE}}=\frac{2\hbar\omega_{\mathrm{e}}Q_{\mathrm{s}}}%
{\mu_{0}V_{\mathrm{s}}M_{\mathrm{s}}}\mathbf{\hat{q}}=\frac{\omega
_{\mathrm{e}}Q_{\mathrm{s}}}{\mu_{0}\gamma_{\mathrm{e}}}\theta_{\mathrm{mz}%
}\mathbf{\hat{q}}\;.\label{H_IFE}%
\end{equation}
Note that the above result (\ref{H_IFE}), which is based on a semiclassical
model \cite{Kusminskiy_299,Zhu_2012_11119}, was found to underestimate the
experimentally measured $H_{\mathrm{IFE}}$ by several orders of magnitudes
\cite{Hansteen_014421,Mikhaylovskiy_100405}. A photon-magnon scattering model
is employed in \cite{rezende2020fundamentals,Fleury_514,Sandercock_1729} to
evaluate $\mathbf{H}_{\mathrm{IFE}}$. For a single photon excitation
$\omega_{\mathrm{e}}=2\pi c/\lambda$, where $\lambda$ is the optical
wavelength, and the corresponding rotation angle of the magnetization, which
is denoted by $\theta_{\mathrm{IFE}}$, is given by [see Eq. (\ref{H_IFE})]]%
\begin{equation}
\theta_{\mathrm{IFE}}=\mu_{0}\gamma_{\mathrm{e}}H_{\mathrm{IFE}}\times
\frac{2n_{0}R_{\mathrm{s}}}{c}\;,\label{theta_IFE}%
\end{equation}
hence \ $\theta_{\mathrm{IFE}}=4\pi n_{0}Q_{\mathrm{s}}R_{\mathrm{s}}%
\theta_{\mathrm{mz}}/\lambda$, or $\theta_{\mathrm{IFE}}=0.18\left(
n_{0}/2.19\right)  \left(  Q_{\mathrm{s}}/10^{-4}\right)  \left(
R_{\mathrm{s}}/100%
%TCIMACRO{\unit{\U{3bc}m}}%
%BeginExpansion
\operatorname{\mu m}%
%EndExpansion
\right)  \left(  \lambda/1550%
%TCIMACRO{\unit{nm}}%
%BeginExpansion
\operatorname{nm}%
%EndExpansion
\right)  ^{-1}\theta_{\mathrm{mz}}$.

\newpage
\bibliographystyle{ieeepes}
\bibliography{acompat,Eyal_Bib}

\end{document}